\documentclass[prd,preprint,nofootinbib]{revtex4}

\begin{document}
%%%%%%%%%%%%%%%%%%%%%%%%%%%%%%%%%

\renewcommand{\thefootnote}{\fnsymbol{footnote}}

\preprint{DESY 07-129}
\title{Suppression of Supergravity Anomalies in Conformal Sequestering}

\renewcommand{\thefootnote}{\alph{footnote}}

\author{Motoi Endo}

\affiliation{
  Deutsches Elektronen Synchrotron DESY, 
  Notkestrasse 85,
  22607 Hamburg, Germany
}

\begin{abstract}
\noindent
We show that the anomaly-mediated supersymmetry breaking 
via the K\"ahler and sigma-model anomalies is suppressed 
by conformal dynamics in the supersymmetry breaking sector. 
\end{abstract}

\maketitle

%%%%%%%%%%%%%%%%%%%%%%%%%%%%%%%%%%%%
\section{Introduction}
%%%%%%%%%%%%%%%%%%%%%%%%%%%%%%%%%%%%

Low-energy supersymmetry (SUSY) is one of 
the most plausible extensions of the standard 
model (SM). So far, low-energy experiments 
such as measurements of flavor-changing neutral 
currents (FCNCs) have imposed constraints on 
its breaking mechanism and mediation. We often 
assume to put our world be secluded from the 
SUSY breaking sector. Then, the SUSY breaking 
is mediated only via the gravitational 
effects~\cite{Randall:1998uk,Giudice:1998xp,Bagger:1999rd}, 
and the dangerous FCNCs are suppressed naturally. 

It was proposed that the separation is achieved 
by geometrical configuration in higher 
dimensions~\cite{Randall:1998uk}. This mechanism 
is simple and easy to imagine. However, it has 
been noted that moduli fields in the bulk may 
induce the dangerous couplings. The contributions 
depend on the background, and the warped one, 
namely the AdS space, is successful, because 
they are warped away~\cite{Luty:2000ec}. 

On the other hand, the separation is realized 
in the four dimensional setup by assuming a 
conformal dynamics in the SUSY breaking sector. 
This scenario is called as the conformal 
sequestering~\cite{Luty:2001zv}. 
The renormalization group (RG) evolution of 
the conformal dynamics suppresses the contact 
couplings between the SM and SUSY breaking 
sectors. 

These two mechanisms are suggested to be 
dual to each other according to the AdS/conformal 
field theory (CFT) correspondence~\cite{Maldacena:1997re}. 
This implies an equivalence of the mass 
spectrum of the superparticles. It has been 
studied that the tree-level mediation 
of the SUSY breaking is suppressed in both 
cases~\cite{Randall:1998uk,Luty:2001zv}. 
Then the soft parameters arise at the quantum 
level. There are three anomalies in supergravity 
(SUGRA), which are known to mediate the SUSY 
breaking~\cite{Randall:1998uk,Giudice:1998xp,Bagger:1999rd}. 
In the AdS setup, the mediation is given by 
the Super-Weyl (SW) anomaly, while the other 
two anomalies in SUGRA, called the K\"ahler 
and sigma-model anomalies, are known to 
cancel to each other~\cite{Bagger:1999rd}. 
In contrast, any cancellation or suppression 
has not been discussed in CFT. In this letter, 
we will show that the conformal dynamics 
suppresses the K\"ahler and sigma-model 
anomalies are suppressed.

%%%%%%%%%%%%%%%%%%%%%%%%%%%%%%%%%%%%
\section{Anomaly Mediation}
%%%%%%%%%%%%%%%%%%%%%%%%%%%%%%%%%%%%

The anomaly-mediated SUSY breaking (AMSB) with 
respect to the SW, K\"ahler and sigma-model 
transformations is represented by the non-local 
operators in SUGRA~\cite{Bagger:1999rd}. 
However, the result is not easy to discuss 
the conformal dynamics. They are easily 
obtained from the superconformal formula of 
SUGRA~\cite{Cremmer:1978hn}. Only the leading 
terms with respect to $1/M_P$ are phenomenologically 
significant. Then the Lagrangian is expanded as
\begin{eqnarray}
    \mathcal{L} \;=\; 
    [\phi^\dagger\phi\, Q^\dagger Q]_D + 
    [\Delta K]_D - 
    \frac{1}{6} [K^2]_D +
    [\phi^3\, W]_F + \cdots,
    \label{eq:Lagrangian}
\end{eqnarray}
where $K$ and $W$ denotes the K\"ahler and 
superpotential in the Einstein frame. The 
chiral superfield field $Q$ denote the visible 
and hidden mattes. It is noted that $\phi$ 
is the chiral compensator field to fix the 
gauge degrees of freedom of the superconformal 
symmetry. Namely, the frame is not fixed 
before giving a VEV for $\phi$. The notation 
$[\cdots]_{D,F}$ means to take $D$- and 
$F$-components in the global SUSY, respectively. 
Further, we simply assume a canonical 
normalization for the matters. The second 
term in the right-handed side represents 
the higher dimensional terms, potentially 
including direct couplings between the visible 
and hidden sectors. The third one is obtained 
after expanding $-3 e^{-K/3}$. The neglected 
terms are phenomenologically irrelevant, 
since they correspond to higher order terms 
of $1/M_P^n$ in the Einstein frame. 

The chiral compensator field, $\phi$ is a 
source to mediate the SUSY breaking via 
the SW anomaly. It is easy to introduce the 
Pauli-Villas (PV) fields $Q'$ to see AMSB. 
Essentially, the superpotential involves 
the mass term, 
\begin{eqnarray}
    W \;=\; M' Q' \bar Q'
    \label{eq:PV-mass}
\end{eqnarray}
with the regularization scale $M'$. After 
canonically rescaling $Q'$, the SUSY 
breaking B term is evaluated as $B = M' 
F_\phi$ in addition to the mass term $M = M' 
\phi$. Thus similarly to the evaluation 
of the gaugino mass in the gauge-mediated 
SUSY breaking, the loop diagram mediating 
$Q'$ gives 
\begin{eqnarray}
    M_\lambda \;=\; 
    \frac{\alpha}{4\pi} \frac{F_\phi}{\phi}.
    \label{eq:SW-AMSB}
\end{eqnarray}
This has a sign opposite to that of the 
gauge-mediation because $Q'$ is the PV 
field. We notice that the result is 
independent of $M'$ and finite even for 
$M' \to \infty$. The Einstein frame is 
realized by taking~\footnote{
  See \cite{Bagger:1999rd} for the terms 
  involving spinors.
} 
\begin{eqnarray}
    \phi \;=\; e^{K/6} 
    \left[1 + \theta^2 
      \left( 
        e^{K/2} W^* + 
        \frac{1}{3} K_i F^i 
      \right) 
    \right].
    \label{eq:Einstein-frame}
\end{eqnarray}
Then we  reproduce the AMSB result from the 
SW anomaly. 

The sigma-model contribution originates in 
the second term of the right-handed side 
in (\ref{eq:Lagrangian}). The B term is from 
the higher dimensional operator in the K\"ahler 
potential. In fact, for a hidden matter $Z$, 
$\delta K = c Z Q' Q^{\prime \dagger} + 
{\rm h.c.}$ gives $\delta F_{Q'} = - c F_Z Q'$, 
leading to $B = - M' c F_Z$ by combining 
to the mass term (\ref{eq:PV-mass}) 
(e.g. see below). Note that $\phi$ does not 
contribute to the sigma-model anomaly. 
Thus the gaugino mass becomes
\begin{eqnarray}
    M_\lambda \;=\; 
    - \frac{\alpha}{4\pi} c F_Z.
    \label{eq:sm-AMSB}
\end{eqnarray}
This result is generalized to the result 
in \cite{Bagger:1999rd} straight-forwardly. 
Then the anomaly is only from the $U(1)$ 
subgroup of the connection, $\Gamma^j_{ij} 
\equiv K^{j\ell^*} K_{i\ell^* j}$. It is also 
commented that this result depends on the 
higher dimensional operator in $K$ and can 
appear in global SUSY models~\cite{Dine:2007me}. 

Let us discuss the K\"ahler anomaly. The 
third term of the right-handed side plays 
a role to mediate the SUSY breaking in 
(\ref{eq:Lagrangian}). It looks like a 
higher dimensional operator in the $D$-term, 
$[\cdots]_D$, and the B term becomes $B = 
2/3 M' K_Z F_Z$ for both $Q$ and $\bar Q$, 
similarly to the sigma-model anomaly. 
So the gaugino mass is 
\begin{eqnarray}
    M_\lambda \;=\; 
    \frac{\alpha}{4\pi} \frac{2}{3} K_Z F_Z.
    \label{eq:K-AMSB}
\end{eqnarray}
It is stressed that although the result 
depends on the linear term of $K$, it 
substantially comes from the higher 
dimensional operator in (\ref{eq:Lagrangian}). 

From (\ref{eq:SW-AMSB}), (\ref{eq:sm-AMSB}) 
and (\ref{eq:K-AMSB}), we obtain the complete 
AMSB for the gaugino mass which is coincide 
with the result in \cite{Bagger:1999rd}. 
In the literature, the operator is denoted 
by the superfields, involving the gravity 
superfield, $R$. We can see that the superfield 
representation of the non-local terms is 
derived from the second and third terms in 
(\ref{eq:Lagrangian}) for the K\"ahler and 
sigma-model anomalies. However, only a part 
is obtained for that of the SW anomaly, 
because we focus on a source of AMSB and 
introduced only $\phi$ in this letter. 

The B terms are essential to derive AMSB 
in the above. For the K\"ahler and sigma-model 
anomalies, they come from the higher dimensional 
operators. The K\"ahler potential is generally 
written as (here and in the following, we omit 
a prime of fields for simplicity)
\begin{eqnarray}
    K \;=\; |Z|^2 + |Q|^2 + |\bar Q|^2 + 
    \left[ d\, Z + c_Q Z |Q|^2 + 
      c_{\bar Q} Z|\bar Q|^2 + {\rm h.c.} 
    \right] + \cdots,
    \label{eq:general-K}
\end{eqnarray}
and the mass term is $W = M Q \bar Q$. 
Here the coefficients $c_{Q,\bar Q},\,d$ 
may depend on the (hidden) matters as a 
background. Expanding $e^{K/3}$, we obtain 
the higher dimensional operators;
\begin{eqnarray}
    -3 e^{-K/3} \;\supset\;
    ( c_Q - d/3 )\, Z |Q|^2 + 
    ( c_{\bar Q} - d/3 )\, Z |\bar Q|^2 + 
    {\rm h.c.}.
    \label{eq:higher-dim}
\end{eqnarray}
These terms are a source of mediating 
the SUSY breaking in the K\"ahler and 
sigma-model AMSB. The B term is easily 
obtained by solving the equation of 
motion of $F_Q$ and $F_{\bar Q}$. Another 
approach is to erase them by rescaling, 
$Q \to Q [ 1 - ( c_Q - d/3 ) Z ]$. Then 
the mass term is modified as
\begin{eqnarray}
    M Q \bar Q \;\longrightarrow\; 
    M \left[ 1 - \left( 
        c_Q - c_{\bar Q} - \frac{2d}{3} 
      \right) Z \right] 
    Q \bar Q.
    \label{eq:Dterm-K-S}
\end{eqnarray}
This involves the B term, and provides 
the gaugino masses. It is noted that 
the tadpole terms of $Z$ are irrelevant 
after the expansion. 

The contributions from the K\"ahler and 
sigma-model anomalies, (\ref{eq:sm-AMSB}) 
and (\ref{eq:K-AMSB}), exactly cancel to 
each other, if the K\"ahler potential is 
the sequestered form~\cite{Bagger:1999rd}, 
\begin{eqnarray}
    K \;=\; -3 \ln 
    \left[ 
      1 - \frac{1}{3} (|Q|^2 + |Z|^2) 
    \right]. 
    \label{eq:sequesterd-form}
\end{eqnarray}
This cancellation is easily seen in 
(\ref{eq:Lagrangian}). The second and third 
terms in the right-handed side are a source 
of the SUSY breaking for the sigma-model 
and K\"ahler anomalies. If we substitute 
(\ref{eq:sequesterd-form}) for the K\"ahler 
potential in (\ref{eq:Lagrangian}), they 
cancel to each other. From another point 
of view, they correspond to the higher 
dimensional operators of $-3 e^{K/3}$. 
Namely, the higher dimensional operators 
in the Einstein frame are practically 
equivalent to those in the conformal 
frame~\cite{Endo:2007sz}. In the conformal 
frame, since (\ref{eq:sequesterd-form}) 
does not have the contact terms between 
the visible and the SUSY breaking sectors, 
the K\"ahler and sigma-model anomalies 
are absent, and only the SW anomaly remains.

%%%%%%%%%%%%%%%%%%%%%%%%%%%%%%%%%%%%
\section{Conformal Sequestering}
%%%%%%%%%%%%%%%%%%%%%%%%%%%%%%%%%%%%

Let us discuss the K\"ahler and sigma-model 
anomalies under the conformal dynamics. 
In the previous section, we saw that they 
are related to the higher dimensional 
operators in (\ref{eq:Lagrangian}). Thus 
we focus on the evolution of them in the 
conformal dynamics. 

At the cutoff scale, the Lagrangian is 
assumed to be general, involving the 
(flavor-violating) higher dimensional 
operators. Let us first discuss the case 
when the operators in the $D$-term linearly 
depend on the matters in the SUSY breaking 
sector, $S$. This means that $c_{Q,\bar Q}$ 
and $d$ in (\ref{eq:general-K}) are 
independent of the SUSY breaking fields. 
To see a suppression of them, we rescale 
the visible matters as $Q \to Q [ 1 - 
( c_Q - d/3 ) S ]$. Then the AM contributions 
is derived from a coupling of $S$ in front 
of the mass term in the superpotential, 
giving the B term. Its evolution is 
represented by the anomalous dimension 
of $S$. Near the fixed point, the B term 
behaves as (see e.g. \cite{Schmaltz:2006qs,Murayama:2007ge}) 
\begin{eqnarray}
    W \;\sim\; 
    \left( 
      \frac{\mu}{M_*} 
    \right)^{\gamma^*_S}
    M S Q \bar Q,
\end{eqnarray}
where $\gamma^*_S$ is the anomalous 
dimension at the fixed point. Since 
$S$ should be gauge-singlet, $\gamma^*_S$ 
is positive. Thus the B term becomes 
suppressed in the infrared limit. 

The bilinear terms with respect to the 
SUSY breaking fields in the $D$-term can 
also be a source of mediating the SUSY 
breaking if the field has a finite vacuum 
expectation value. Regarding the visible 
fields as a background, their evolutions 
are represented by the anomalous 
dimensions~\cite{Luty:2001zv}; 
\begin{eqnarray}
    (\Delta \ln Z) = e^{Lt} (\Delta \ln Z)_0.
\end{eqnarray}
Here the scale is $t = \ln (\mu/M_*)$ 
and $(\Delta \ln Z)$ is defined as 
$(\Delta \ln Z) \equiv \ln Z + \gamma^* t$. 
Since the SUSY breaking sector usually 
consists of multiple fields, $L$ forms 
a matrix. If it is positive, i.e. all 
eigenvalues are positive, $(\Delta \ln Z)$ 
approaches to zero for the infrared 
limit $t \to - \infty$. Then the 
contact terms are absent from the 
low-energy effective Lagrangian, 
because they arise as $(\Delta \ln Z)_0 
\supset c Q Q^\dagger$. Therefore the 
conformal sequestering is realized for 
$L > 0$~\cite{Luty:2001zv,Schmaltz:2006qs,Ibe:2005pj}. 
At the same time, the sources of the 
SUSY breaking mediation become small 
as well, because they are denoted by the 
higher dimensional operators. Thus the 
K\"ahler and sigma-model anomalies are 
suppressed by the conformal dynamics. 
Although the coefficients $c$ and $d$ 
in (\ref{eq:Dterm-K-S}) may depends on 
the hidden matters more complexly, they 
can be treated similarly, or are practically 
irrelevant for phenomenology. 

Consequently, the B terms relevant for 
the K\"ahler and sigma-model anomalies 
are suppressed, and so they are absent 
in the conformal sequestering. 
In contrast, the SW anomaly still 
remains after the dynamics, since $\phi$ 
arises as an overall factor in front of 
the $D$-term~\footnote{
  The conformal dynamics may affect 
  $K_iF^i/3$ in (\ref{eq:Einstein-frame}). 
  The evolution, however, depends on 
  details of SUGRA, and we retain the 
  discussion for a future work. 
}. 

Let us comment on a choice of the 
regularization scheme. So far, we used the 
PV regularization. If we apply the other 
scheme (see e.g. \cite{Bagger:1999rd,Dine:2007me,Boyda:2001nh}), 
the discussions in the above are not so 
trivial. In order to see the suppression 
of AMSB, we focus on the UV insensitivity. 
When a matter field decouples by a heavy 
mass, the threshold corrections give the 
gaugino mass, $M_\lambda^{({\rm dec.})}$. 
The UV insensitivity tells us that it exactly 
cancels with that from the regularization, 
that is, the AMSB mass, $M_\lambda^{({\rm AM})}$. 
Thus if we evaluate the gaugino mass from 
the matter threshold by postulating a 
hypothetical mass term, we obtain the AMSB 
mass as $M_\lambda^{({\rm AM})} = - M_\lambda^{({\rm dec.})}$. 
Repeating the same discussions in this 
letter, we obtain the same result. 

So far, we focused on the gaugino mass. 
The soft SUSY breaking effects also contain 
scalar masses, scalar trilinear couplings, 
and holomorphic scalar mass terms. The 
SUGRA anomalies mediate the SUSY breaking 
to the parameters. Nevertheless, the complete 
result has not been known for the K\"ahler 
and sigma-model anomalies (see also 
\cite{Gaillard:2000fk}). On the other hand, 
the SUSY breaking is mediated by the higher 
dimensional operators in (\ref{eq:Lagrangian}). 
The soft parameters other than the gaugino 
mass are also considered to originate in 
the terms. We saw that they are suppressed 
in the geometrical and conformal sequestering. 
Thus, if the sequestering is realized in nature, 
the K\"ahler and sigma-model anomalies do not 
contribute to the soft parameters.

%%%%%%%%%%%%%%%%%%%%%%%%%%%%%%%%%%%%
\section{Discussion and Conclusions}
%%%%%%%%%%%%%%%%%%%%%%%%%%%%%%%%%%%%

In this letter, we discuss the suppression 
of the K\"ahler and sigma-model anomalies 
in the conformal sequestering. The contributions 
are obtained from the higher dimensional 
operators in the $D$-term, namely after 
expanding $-3 e^{-K/3}$. Since the conformal 
dynamics suppresses them, the anomalies are 
found to vanish. 

A dynamics of the gauge term $\int d^2\theta 
Z WW$ is treated by using the anomalous 
dimensions~\cite{Murayama:2007ge}. However, 
the operators we focus on now are represented 
by the non-local operators at the Planck 
scale~\cite{Bagger:1999rd}, so its evolution 
is non-trivial. Instead, the counter term 
may exist at the cutoff, and can affect the 
gaugino mass~\cite{Bagger:1999rd}. If it has 
a form of $\int d^2\theta f(Z) WW$, where 
$f(Z) = \alpha Z + \cdots$ is a function of 
$Z$, its contribution tends to be suppressed 
by the conformal dynamics. 

The method in this letter can also be applied 
to discuss the anomaly-induced inflaton 
decay~\cite{Endo:2007ih,Endo:2007sz}. The decay 
into the SUSY breaking sector is obtained 
by the higher dimensional operators of $Z$ 
in the $D$-term for the K\"ahler and sigma-model 
anomalies. Thus they are naturally suppressed 
by the conformal dynamics, even when the SUSY 
breaking fields do not always appear explicitly 
in the operators~\cite{ETY}.

%%%%%%%%%%%%%%%%%%%%%%%%%%%%%%%%%%%%
\section*{Acknowledgment}
%%%%%%%%%%%%%%%%%%%%%%%%%%%%%%%%%%%%

The author is grateful to K.-I.~Izawa for fruitful 
discussions. 

%%%%%%%%%%%%%%%%%%%%%%%%%%%%%%%%%%%%

\end{document}